\documentstyle[12pt]{article}
\makeatletter
\newdimen\normalarrayskip              
\newdimen\minarrayskip                 
\normalarrayskip\baselineskip
\minarrayskip\jot
\newif\ifold             \oldfalse
\newif\ifdisplayarray    \displayarraytrue
\newif\ifbigarray        \bigarraytrue
\def\arraymode{\ifold\relax\else\ifdisplayarray\displaystyle\else\relax\fi\fi}
\def\eqnumphantom{\phantom{(\theequation)}}
\def\@arrayskip{\ifold\baselineskip\z@\lineskip\z@\else\ifbigarray
     \baselineskip\normalarrayskip\lineskip\minarrayskip
     \else
     \baselineskip\z@\lineskip\z@\fi\fi}
\def\@arrayclassz{\ifcase \@lastchclass \@acolampacol \or
\@ampacol \or \or \or \@addamp \or
   \@acolampacol \or \@firstampfalse \@acol \fi
\edef\@preamble{\@preamble
  \ifcase \@chnum
     \hfil$\relax\arraymode\@sharp$\hfil
     \or $\relax\arraymode\@sharp$\hfil
     \or \hfil$\relax\arraymode\@sharp$\fi}}
\def\@array[#1]#2{\setbox\@arstrutbox=\hbox{\vrule
     height\arraystretch \ht\strutbox
     depth\arraystretch \dp\strutbox
     width\z@}\@mkpream{#2}\edef\@preamble{\halign \noexpand\@halignto
\bgroup \tabskip\z@ \@arstrut \@preamble \tabskip\z@ \cr}%
\let\@startpbox\@@startpbox \let\@endpbox\@@endpbox
  \if #1t\vtop \else \if#1b\vbox \else \vcenter \fi\fi
  \bgroup \let\par\relax
  \let\@sharp##\let\protect\relax
  \@arrayskip\@preamble}
\def\eqnarray{\stepcounter{equation}%
              \let\@currentlabel=\theequation
              \global\@eqnswtrue
              \global\@eqcnt\z@
              \tabskip\@centering
              \let\\=\@eqncr
              $$%
 \halign to \displaywidth\bgroup
    \eqnumphantom\@eqnsel\hskip\@centering
    $\displaystyle \tabskip\z@ {##}$%
    &\global\@eqcnt\@ne \hskip 2\arraycolsep
         \hfil$\arraymode{##}$\hfil
    &\global\@eqcnt\tw@ \hskip 2\arraycolsep
         $\displaystyle\tabskip\z@{##}$\hfil
         \tabskip\@centering
    &{##}\tabskip\z@\cr}
\newenvironment{marray}{\begin{equation}\begin{array}}%
{\end{array}\end{equation}}
\newenvironment{carray}{\begin{equation}\begin{array}{rcl}}%
{\end{array}\end{equation}}
\def\be{\@ifnextchar[{\def\ee{\end{equation}}\begin{equation}\l@b}%
{\def\ee{$$}$$}}
\def\l@b[#1]{\label{#1}}
\def\ba{\@ifnextchar[{\def\ee{\end{carray}}\begin{carray}\l@b}%
{\def\ee{\end{array}$$}$$\begin{array}{rcl}}}
\def\barray#1{\@ifnextchar[{\def\ee{\end{marray}}\begin{marray}{#1}\l@b}%
{\def\ee{\end{array}$$}$$\begin{array}{#1}}}
\def\herring{\@ifnextchar[{\@herring}{\@herring[\vcenter]}}
\def\@herring[#1]#2{\begingroup
\def\*{\\ \>}
\topsep0pt
\partopsep0pt
\def\tabbing{\lineskip\jot \lineskiplimit\jot
     \let\>\@rtab\let\<\@ltab\let\=\@settab
     \let\+\@tabplus\let\-\@tabminus\let\`\@tabrj\let\'\@tablab
     \let\\=\@tabcr
     \global\@hightab\@firsttab
     \global\@nxttabmar\@firsttab
     \dimen\@firsttab\@totalleftmargin
     \global\@tabpush0 \global\@rjfieldfalse
     \trivlist \item[]\if@minipage\else\vskip\parskip\fi
     \setbox\@tabfbox\hbox{\rlap{\indent\hskip\@totalleftmargin
       \the\everypar}}\def\@itemfudge{\box\@tabfbox}\@startline\ignorespaces}
\def\@startfield{\global\setbox\@curfield\hbox
                    \bgroup$\displaystyle}%
\def\@stopfield{$\egroup}%
#1{\begin{tabbing}#2\end{tabbing}}\endgroup}

\def\eq#1{(\ref{#1})}
\def\theequation{\thesection.\arabic{equation}}
\@addtoreset{equation}{section}%
\def\@cite#1#2{\hbox{ [#1\if@tempswa ,#2\fi]}}
\def\@citex[#1]#2{\if@filesw\immediate\write\@auxout{\string\citation{#2}}\fi
  \def\@citea{}\@cite{\@for\@citeb:=#2\do
    {\@citea\def\@citea{,\penalty\@m}\@ifundefined  
       {b@\@citeb}{{\bf ?}\@warning
       {Citation `\@citeb' on page \thepage \space undefined}}%
\hbox{\csname b@\@citeb\endcsname}}}{#1}}
\def\@sect#1#2#3#4#5#6[#7]#8{\ifnum #2>\c@secnumdepth
     \def\@svsec{}\else
     \refstepcounter{#1}\edef\@svsec{\csname the#1\endcsname.%
     \hskip 0.8em }\fi
     \@tempskipa #5\relax
      \ifdim \@tempskipa>\z@
        \begingroup #6\relax
          \@hangfrom{\hskip #3\relax\@svsec}{\interlinepenalty \@M #8\par}%
        \endgroup
       \csname #1mark\endcsname{#7}\addcontentsline
         {toc}{#1}{\ifnum #2>\c@secnumdepth \else
                      \protect\numberline{\csname the#1\endcsname}\fi
                    #7}\else
        \def\@svsechd{#6\hskip #3\@svsec #8\csname #1mark\endcsname
                      {#7}\addcontentsline
                           {toc}{#1}{\ifnum #2>\c@secnumdepth \else
                             \protect\numberline{\csname the#1\endcsname}\fi
                       #7}}\fi
     \@xsect{#5}}

\makeatother

\begin{document}
\newcommand {\ignore}[1]{}
\newcommand{\nota}[1]{\makebox[0pt]{\,\,\,\,\,/}#1}
\newcommand{\notp}[1]{\makebox[0pt]{\,\,\,\,/}#1}
\newcommand{\braket}[1]{\mbox{$<$}#1\mbox{$>$}}
\newcommand{\Frac}[2]{\frac{\displaystyle #1}{\displaystyle #2}}
\renewcommand{\arraystretch}{1.5}
\newcommand{\noi}{\noindent}
\newcommand{\bc}{\begin{center}}
\newcommand{\ec}{\end{center}}
\newcommand{\epm}{e^+e^-}
\def\ne{\hbox{$\nu_e$ }}
\def\nm{\hbox{$\nu_\mu$ }}
\def\bC{\mathop{\bf C}}
\def\eq#1{{eq. (\ref{#1})}}
\def\Eq#1{{Eq. (\ref{#1})}}
\def\Eqs#1#2{{Eqs. (\ref{#1}) and (\ref{#2})}}
\def\Eqs#1#2#3{{Eqs. (\ref{#1}), (\ref{#2}) and (\ref{#3})}}
\def\Eqs#1#2#3#4{{Eqs. (\ref{#1}), (\ref{#2}), (\ref{#3}) and (\ref{#4})}}
\def\eqs#1#2{{eqs. (\ref{#1}) and (\ref{#2})}}
\def\eqs#1#2#3{{eqs. (\ref{#1}), (\ref{#2}) and (\ref{#3})}}
\def\eqs#1#2#3#4{{eqs. (\ref{#1}), (\ref{#2}), (\ref{#3}) and (\ref{#4})}}
\def\fig#1{{Fig. (\ref{#1})}}
\def\lie{\hbox{\it \$}} 
\def\partder#1#2{{\partial #1\over\partial #2}}
\def\secder#1#2#3{{\partial^2 #1\over\partial #2 \partial #3}}
\def\bra#1{\left\langle #1\right|}
\def\ket#1{\left| #1\right\rangle}
\def\VEV#1{\left\langle #1\right\rangle}
\let\vev\VEV
\def\gdot#1{\rlap{$#1$}/}
\def\abs#1{\left| #1\right|}
\def\pri#1{#1^\prime}
\def\ltap{\raisebox{-.4ex}{\rlap{$\sim$}} \raisebox{.4ex}{$<$}}
\def\gtap{\raisebox{-.4ex}{\rlap{$\sim$}} \raisebox{.4ex}{$>$}}
\def\lsim{\raise0.3ex\hbox{$\;<$\kern-0.75em\raise-1.1ex\hbox{$\sim\;$}}}
\def\gsim{\raise0.3ex\hbox{$\;>$\kern-0.75em\raise-1.1ex\hbox{$\sim\;$}}}
\def\contract{\makebox[1.2em][c]{
        \mbox{\rule{.6em}{.01truein}\rule{.01truein}{.6em}}}}
\def\half{{1\over 2}}
\def\bel{\begin{letter}}
\def\eel{\end{letter}}
\def\beq{\begin{equation}}
\def\eeq{\end{equation}}
\def\bef{\begin{figure}}
\def\eef{\end{figure}}
\def\bet{\begin{table}}
\def\eet{\end{table}}
\def\bea{\begin{eqnarray}}
\def\ba{\begin{array}}
\def\ea{\end{array}}
\def\bi{\begin{itemize}}
\def\ei{\end{itemize}}
\def\ben{\begin{enumerate}}
\def\een{\end{enumerate}}
\def\ra{\rightarrow}
\def\ot{\otimes}
\def\lrover#1{
        \raisebox{1.3ex}{\rlap{$\leftrightarrow$}} \raisebox{ 0ex}{$#1$}}
\def\com#1#2{
        \left[#1, #2\right]}
\def\eea{\end{eqnarray}}
\def\bentarrow{\:\raisebox{1.3ex}{\rlap{$\vert$}}\!\rightarrow}
\def\longbent{\:\raisebox{3.5ex}{\rlap{$\vert$}}\raisebox{1.3ex}%
        {\rlap{$\vert$}}\!\rightarrow}
\def\onedk#1#2{
        \begin{equation}
        \begin{array}{l}
         #1 \\
         \bentarrow #2
        \end{array}
        \end{equation}
                }
\def\dk#1#2#3{
        \begin{equation}
        \begin{array}{r c l}
        #1 & \rightarrow & #2 \\
         & & \bentarrow #3
        \end{array}
        \end{equation}
                }
\def\dkp#1#2#3#4{
        \begin{equation}
        \begin{array}{r c l}
        #1 & \rightarrow & #2#3 \\
         & & \phantom{\; #2}\bentarrow #4
        \end{array}
        \end{equation}
                }
\def\bothdk#1#2#3#4#5{
        \begin{equation}
        \begin{array}{r c l}
        #1 & \rightarrow & #2#3 \\
         & & \:\raisebox{1.3ex}{\rlap{$\vert$}}\raisebox{-0.5ex}{$\vert$}%
        \phantom{#2}\!\bentarrow #4 \\
         & & \bentarrow #5
        \end{array}
        \end{equation}
                }
\def\ap#1#2#3{           {\it Ann. Phys. (NY) }{\bf #1} (19#2) #3}
\def\arnps#1#2#3{        {\it Ann. Rev. Nucl. Part. Sci. }{\bf #1} (19#2) #3}
\def\cnpp#1#2#3{        {\it Comm. Nucl. Part. Phys. }{\bf #1} (19#2) #3}
\def\apj#1#2#3{          {\it Astrophys. J. }{\bf #1} (19#2) #3}
\def\app#1#2#3{          {\it Astropart. Phys. }{\bf #1} (19#2) #3}
\def\asr#1#2#3{          {\it Astrophys. Space Rev. }{\bf #1} (19#2) #3}
\def\ass#1#2#3{          {\it Astrophys. Space Sci. }{\bf #1} (19#2) #3}
\def\aa#1#2#3{          {\it Astron. \& Astrophys.  }{\bf #1} (19#2) #3}
\def\apjl#1#2#3{         {\it Astrophys. J. Lett. }{\bf #1} (19#2) #3}
\def\ap#1#2#3{         {\it Astropart. Phys. }{\bf #1} (19#2) #3}
\def\ass#1#2#3{          {\it Astrophys. Space Sci. }{\bf #1} (19#2) #3}
\def\jel#1#2#3{         {\it Journal Europhys. Lett. }{\bf #1} (19#2) #3}
\def\ib#1#2#3{           {\it ibid. }{\bf #1} (19#2) #3}
\def\nat#1#2#3{          {\it Nature }{\bf #1} (19#2) #3}
\def\nps#1#2#3{          {\it Nucl. Phys. B (Proc. Suppl.) }
                         {\bf #1} (19#2) #3}
\def\np#1#2#3{           {\it Nucl. Phys. }{\bf #1} (19#2) #3}
\def\pl#1#2#3{           {\it Phys. Lett. }{\bf #1} (19#2) #3}
\def\pr#1#2#3{           {\it Phys. Rev. }{\bf #1} (19#2) #3}
\def\prep#1#2#3{         {\it Phys. Rep. }{\bf #1} (19#2) #3}
\def\prl#1#2#3{          {\it Phys. Rev. Lett. }{\bf #1} (19#2) #3}
\def\pw#1#2#3{          {\it Particle World }{\bf #1} (19#2) #3}
\def\ptp#1#2#3{          {\it Prog. Theor. Phys. }{\bf #1} (19#2) #3}
\def\jppnp#1#2#3{         {\it J. Prog. Part. Nucl. Phys. }{\bf #1} (19#2) #3}
\def\rpp#1#2#3{         {\it Rep. on Prog. in Phys. }{\bf #1} (19#2) #3}
\def\ptps#1#2#3{         {\it Prog. Theor. Phys. Suppl. }{\bf #1} (19#2) #3}
\def\rmp#1#2#3{          {\it Rev. Mod. Phys. }{\bf #1} (19#2) #3}
\def\zp#1#2#3{           {\it Zeit. fur Physik }{\bf #1} (19#2) #3}
\def\fp#1#2#3{           {\it Fortschr. Phys. }{\bf #1} (19#2) #3}
\def\Zp#1#2#3{           {\it Z. Physik }{\bf #1} (19#2) #3}
\def\Sci#1#2#3{          {\it Science }{\bf #1} (19#2) #3}
\def\n.c.#1#2#3{         {\it Nuovo Cim. }{\bf #1} (19#2) #3}
\def\r.n.c.#1#2#3{       {\it Riv. del Nuovo Cim. }{\bf #1} (19#2) #3}
\def\sjnp#1#2#3{         {\it Sov. J. Nucl. Phys. }{\bf #1} (19#2) #3}
\def\yf#1#2#3{           {\it Yad. Fiz. }{\bf #1} (19#2) #3}
\def\zetf#1#2#3{         {\it Z. Eksp. Teor. Fiz. }{\bf #1} (19#2) #3}
\def\zetfpr#1#2#3{         {\it Z. Eksp. Teor. Fiz. Pisma. Red. }{\bf #1}
(19#2) #3}
\def\jetp#1#2#3{         {\it JETP }{\bf #1} (19#2) #3}
\def\mpl#1#2#3{          {\it Mod. Phys. Lett. }{\bf #1} (19#2) #3}
\def\ufn#1#2#3{          {\it Usp. Fiz. Naut. }{\bf #1} (19#2) #3}
\def\sp#1#2#3{           {\it Sov. Phys.-Usp.}{\bf #1} (19#2) #3}
\def\ppnp#1#2#3{           {\it Prog. Part. Nucl. Phys. }{\bf #1} (19#2) #3}
\def\cnpp#1#2#3{           {\it Comm. Nucl. Part. Phys. }{\bf #1} (19#2) #3}
\def\ijmp#1#2#3{           {\it Int. J. Mod. Phys. }{\bf #1} (19#2) #3}
\def\ic#1#2#3{           {\it Investigaci\'on y Ciencia }{\bf #1} (19#2) #3}
\def\tp{these proceedings}
\def\pc{private communication}
\def\opc{\hbox{{\sl op. cit.} }}
\def\ip{in preparation}
\relax
\topmargin -2cm
\textwidth 17cm
\textheight 25.5cm
\evensidemargin  0cm
\def\e{\mbox{e}}
\def\sgn{{\rm sgn}}
\def\gsim{\;
\raise0.3ex\hbox{$>$\kern-0.75em\raise-1.1ex\hbox{$\sim$}}\;
}
\def\lsim{\;
\raise0.3ex\hbox{$<$\kern-0.75em\raise-1.1ex\hbox{$\sim$}}\;
}
\def\MeV{\rm MeV}
\def\eV{\rm eV}
\thispagestyle{empty}
\begin{titlepage}
\begin{center}
\hfill FTUV/95-39\\
\hfill IFIC/95-41\\
\vskip 0.3cm
\LARGE
{\bf Bounds on neutrino
transition magnetic moments in random magnetic fields}
\end{center}
\normalsize
\vskip1cm
\begin{center}
{\bf S. Pastor},
{\bf V.B.Semikoz}
\footnote{On leave from the {\sl Institute of the
Terrestrial Magnetism, the Ionosphere and Radio
Wave Propagation of the Russian Academy of Sciences,
IZMIRAN, Troitsk, Moscow region, 142092, Russia}.}
{\bf and J. W. F. Valle}
\footnote{E-mail valle@flamenco.ific.uv.es}\\
\end{center}
\begin{center}
\baselineskip=13pt
{\it Instituto de F\'{\i}sica Corpuscular - C.S.I.C.\\
Departament de F\'{\i}sica Te\`orica, Universitat de Val\`encia\\}
\baselineskip=12pt
{\it 46100 Burjassot, Val\`encia, SPAIN         }\\
\vglue 0.8cm
\end{center}

\begin{abstract}
We consider the conversions of active to sterile Majorana
neutrinos $\nu_{a}$ and $\nu_{s}$, due to neutrino transition
magnetic moments in the presence of random magnetic fields
(r.m.f.) generated at the electroweak phase transition.
{}From a simple Schr\"{o}dinger-type evolution equation,
we derive a stringent constraint on the corresponding
transition magnetic moments and display it as a function
of the domain size and field geometry. For typical parameter
choices one gets limits much stronger than usually derived from
stellar energy loss considerations. These bounds are consistent
with the hypothesis of seeding of galactic magnetic
fields by primordial fields surviving past the
re-combination epoch.
We also obtain a bound on active-sterile neutrino
transition magnetic moments from supernova energy
loss arguments. For r.m.f. strengths in the range
$10^7$ to $10^{12}$ Gauss we obtain limits varying from
$\mu_{as}^{\nu} \lsim 10^{-13}\mu_B$
to $\mu_{as}^{\nu} \lsim 10^{-18}\mu_B$,
again much stronger than in the case
without magnetic fields.

\end{abstract}

\vfill

\end{titlepage}

\newpage

\section{Introduction}

Recently there has been a renewed interest concerning
neutrino propagation in media with random magnetic fields,
both from the point of view of the early universe hot plasma
as well as astrophysics \cite{Zeldo}.
It has been shown that  random magnetic fields can strongly
influence neutrino conversion rates and this could have
important cosmological and astrophysical implications,
especially in the case of conversions involving a light
sterile neutrino $\nu_s$. Indeed, note that the present
hints from solar and atmospheric neutrino observations
\cite{granadasol,atm} as well as from the COBE
data on cosmic background temperature
anisotropies on large scales \cite{cobe,cobe2} indicate that
a light sterile neutrino $\nu_s$ \cite{DARK92,DARK92B}
might exist in nature.

So far the most stringent constraints for the neutrino
mass matrix including a sterile neutrino species, $\nu_s$,
are obtained from the nucleosynthesis bound on the maximum
number of extra neutrino species that
can reach thermal equilibrium before nucleosynthesis and change
the primordially produced helium abundance \cite{Schramm}.
This has been widely discussed in the case of
vanishing magnetic field \cite{Enqvist} and
for the case of a large
random magnetic field that could arise from the
electroweak phase transition and act as seed
for the galactic magnetic fields \cite{SemikozValle}.

The effect of active-sterile neutrino conversions in a
supernova has also been discussed, both in the case where
no magnetic field is present \cite{Maalampi}, as well as
in the presence of random magnetic field as large as
$10^{16}$ Gauss \cite{sergio}, following a suggestion
made in ref. \cite{Dunkan}.

In this paper we focus on the important effect
of relatively small random magnetic fields on the
active sterile neutrino conversion rates when nonzero
Majorana neutrino transition magnetic moments
are taken into account \cite{BFD,LAM}.
We apply this to the case of $\nu_{a}$ to $\nu_{s}$
conversions in the early universe hot plasma, as well
as in a supernova, showing how their effect can place limits
that are substantially more stringent than those
that apply in the absence of a magnetic field.
First we focus on the nucleosynthesis constraint
on active to sterile transition magnetic moments,
and display this constraint as a function of the
domain size and field geometry. For typical choices
we obtain $\mu_{as}^{\nu} \lsim 10^{-15}\mu_B$,
which is much stronger than usually derived from stellar
energy loss considerations neglecting r.m.f.
effects. These bounds are consistent
with the hypothesis of seeding of galactic magnetic
fields by primordial fields generated at the
electroweak phase transition.
We also obtain a bound on active-sterile neutrino
transition magnetic moments from supernova energy
loss arguments.
For modest r.m.f. field strengths in the range
$10^7$ to $10^{12}$ Gauss we obtain limits that
vary from
$\mu_{as}^{\nu} \lsim 10^{-13}\mu_B$
to $\mu_{as}^{\nu} \lsim 10^{-18}\mu_B$,
again much stronger than in the case
without magnetic fields \cite{BarbieriMohapatra}.


\section{Neutrino conversions}
\vskip 0.5cm
Let us consider the Schr\"{o}dinger equation describing a system
of two neutrinos, one active and one sterile, $\nu_a$ and $\nu_s$
in a plasma with a random magnetic field {\bf B}(t). In general
one has a four-dimensional system of equations \cite{BFD}.
We will use the ultra-relativistic limit and will neglect
the corresponding mixing angle, $s=0$, $c =1$, in which case
one may write \cite{Sergio95}
\barray{ll}[initial]
i\frac{d}{dt}\left (\matrix{\nu_a\\\nu_s}\right )
 = \left (\matrix{V_a - \Delta  +  \mu_{eff}B_{\parallel}(t) & \mu
B_{\perp}(t)\\
           \mu B_{\perp}(t) & 0}\right )\left (\matrix{\nu_a\\\nu_s}\right ).
{}~
\ee
Here $V_a$ (for $\nu_a$, $a =e, \mu, \tau$) is
the active neutrino vector interaction potential.
For instance, for electron left-handed neutrinos
propagating in the early universe hot plasma one
has \cite{Notzold},
\be[potential1]
V = 3.45\times 10^{-20}\Bigl (\frac{T}{MeV}\Bigr )^5 MeV,
\ee
while for the case of a supernova one has
\be[potential2]
V\simeq 4\times 10^{-6}\rho_{14}(3Y_e + 4Y_{\nu_e} + 2Y_{\nu_{\mu}} - 1)MeV.
\ee
where $\rho_{14}$ is the core density in units of
$10^{14}$ g/$cm^3$ .

In the case of the muon left-handed neutrino the last abundance factor in
\eq{potential2} is changed to $f(Y) = Y_e - 1 + 4Y_{\nu_{\mu}} +
2Y_{\nu_e}$, where we took into account $\nu$-$\nu$ forward scattering
amplitude in a supernova matter during the main neutrino burst time, about
10-20 seconds.

The term $\mu_{eff}B_{\parallel}$ is produced by the mean axial vector
current of charged leptons in an external magnetic field
$B_{\parallel}$ = {\bf B.q}$/q$.  In the case of
the early universe hot plasma the coefficient $\mu_{eff}$
is given as
\be[momentBBN]
\mu_{eff} =
-12c_A\times 10^{-13}\mu_B\Bigl (\frac{T}{MeV}\Bigr ),
\ee
while for the case of an ultra-relativistic degenerate
electron gas of a supernova one has
\be[momentsupernova]
\mu_{eff} = - 8.6c_A\times 10^{-13}\mu_B
(p_{F_e}/MeV) \: .
\ee
Here  $c_A = \mp 0.5$ is the axial constant (upper sign
for electron neutrino, lower one for $\nu_{\mu,\tau}$).

Note that even though the quantity $\mu_{eff}$ has the dimensions
of a magnetic moment it is not a real magnetic moment
since it is helicity conserving.
Indeed, the additional energy splitting term
obtained in \eq{initial} does not lead to any
helicity change. In contrast, the off-diagonal
entries of the Hamiltonian \eq{initial} involve
the presence of the real neutrino transition
magnetic moment $\mu = \mu_{as}^{\nu}$ as well
as the transversal magnetic field component,
{\bf B}$_\perp$.

Note also that our initial
equation \eq{initial} is quite general, since it applies
to the case of Majorana neutrinos. It remains valid also
for the description of the active to sterile neutrino
conversions in the limit where these form a Dirac neutrino,
in which case $\Delta = (m_2^2 - m_1^2)/2q = 0$ and $\mu$
becomes the usual magnetic moment.

Denoting $\mu B_{\perp}(t)= \tilde{H}_{\perp}(t)$ and using
the auxiliary functions $R = Re(\langle \nu_e^* \nu_s \rangle)$
and $I = Im(\langle \nu_e^* \nu_s \rangle)$, we obtain
from \eq{initial} the standard system of first order differential
equations,
\be
\begin{array}{ll}
&\dot{P}(t) = - 2\tilde{H}_{\perp}(t)I(t),\\
&\dot{I}(t) = [V - \Delta + \mu_{eff}B_{\parallel}(t)]R(t) + \tilde{H}_{\perp}
(t)(2P(t) -1),\\
&\dot{R}(t) = - [V - \Delta + \mu_{eff}B_{\parallel}(t)]I(t),
\end{array}
\ee
from which we derive the integro-differential equation for the neutrino
conversion probability $P_{\nu_e\to \nu_s}(t) \equiv P(t)$
\be[master]
\dot{P}(t) = - 2\int_0^t\tilde{H}_{\perp}(t)\tilde{H}_{\perp}(t_1)\cos
\Bigl (\int_{t_1}^tV(t_2)dt_2\Bigr )[2P(t_1) - 1]dt_1,
\ee
where $\int_{t_1}^tV(t_2)dt_2 = (V - \Delta)(t - t_1) +  \mu_{eff}\int_{t_1}^t
B_{\parallel}(t_2)dt_2$.

For r.m.f. domain sizes much smaller than the typical
neutrino conversion length $l_{conv} \sim \Gamma^{-1}$,
i.e. $L_0\ll l_{conv}$ (see below \eq{damping})
one can average this equation over the random magnetic
field distribution. The averaged probability
$\langle P(t)\rangle = \cal{P}(t)$ must depend
only on even powers of the random field, since
$\langle (B_j(t))^{2n + 1}\rangle=0$.

In what follows we neglect neutrino collisions with
dense matter. One can show that averaging of the
neutrino conversion probability over fast
collisions does not lead to essential change
of our results
\footnote{
Moreover, in \cite{EnqvistRezSemikoz} it was shown
that in an appropriate kinetic approach for the
neutrino spin evolution in uncorrelated
random magnetic fields the final result does not
depend on the collision frequency at all}.

For uncorrelated random magnetic field domains,
\be[correlator]
\langle B_i(\vec{x})B_j(\vec{y})\rangle =  \frac{1}{2\lambda}\delta_{ij}
\delta^{(3)}(\vec{x} - \vec{y}),
\ee
we obtain the mean values
\be[correlators]
\begin{array}{ll}
&\langle B_{\parallel}(t)\rangle = \langle B_{\perp}(t)\rangle =
\langle B_{\parallel}(t)B_{\perp}(t_1)\rangle =0,\\
&\langle B_{\parallel}(t)B_{\parallel}(t_1)\rangle = \langle B_{\parallel}^2
\rangle L_0\delta (t - t_1),\\
&\langle B_{\perp}(t)B_{\perp}(t_1)\rangle = \langle B_{\perp}^2
\rangle L_0\delta (t - t_1),\\
\end{array}
\ee
The correlation length $\lambda$ is determined by the
domain size $L_0$ and by the value of root mean squared
field $B \equiv B_{rms} \equiv \sqrt{\langle B^2\rangle}$
at the horizon scale $L = l_H$ \cite{EnqvistRezSemikoz}:
$$
\frac{1}{\lambda} =\frac{3}{\pi (3 - 2p)}B_{rms}^2(l_H)L_0^3,
$$
for $p \neq 3/2$, and
$$
\frac{1}{\lambda} =\frac{3}{\pi} \ln \frac{l_H}{L_0}B_{rms}^2(l_H)L_0^3,
$$
for $p =3/2$.

In \eq{correlators} the mean squared fields
$\langle B_{\parallel}^2\rangle =
\langle B^2\rangle/3$, $\langle B_{\perp}^2\rangle = 2\langle B^2\rangle/3$
are given by the r.m.s. field $B \equiv \sqrt{\langle B^2\rangle}$.

Averaging the master equation \eq{master} over a random
magnetic field distribution with the use of \eq{correlators}
we easily obtain the simple differential equation
\be[averaged]
\dot{\cal{P}} + \Gamma_{\perp} \cal{P} = \frac{\Gamma_{\perp}}{2},
\ee
which has the solution
\be[solution]
\cal{P} = \frac{1}{2}( 1 - \exp (-\Gamma_{\perp} t))
\approx \frac{\Gamma_{\perp} t}{2},
\ee
where the magnetic field damping parameter $\Gamma_{\perp}$ is
defined as
\be
[damping]
\Gamma_{\perp} = 4\mu^2\langle B^2_{\perp} \rangle L_0 =
\frac{8}{3}\mu^2B^2L_0.
\ee
Note that the result \eq{solution} obtained
from our Schr\"{o}dinger equation \eq{initial} coincides
with the analogous ones obtained in \cite{EnqvistRezSemikoz}
for the Dirac neutrino magnetic moment, using neutrino spin
kinetic equation \cite{Semikoz}, or the Redfield equation for
the density matrix describing a nuclear spin interacting
with the lattice vibrations in solids \cite{BalantekinLoreti}.

Note also that in the simplified case of negligible
neutrino mixing considered here, the assumed $\delta$-correlated
form of the random magnetic field implies that there is no dependence
of the averaged neutrino conversion probability on
the neutrino energy splitting due to the matter term $V$,
to the term $\mu_{eff}B_{\parallel}$ or to the mixing
parameter $\Delta$.

\section{Nucleosynthesis constraint}

The rate for producing sterile neutrinos in the
early universe hot plasma is given as the product
of a typical weak interaction rate $\Gamma_W$ times
our averaged conversion probability $\cal{P}$. The
constraint that follows from nucleosynthesis may
be simply estimated as
\be[bbnconstraint]
\Gamma_s = \Gamma_W \cal{P} \lsim H,
\ee
where $\Gamma_W = 4.0G_F^2T^5$ is the rate for
producing the standard model active neutrinos,
$H = 4.46\times 10^{-22}(T/MeV)^2$$MeV$ is
the Hubble parameter, and $\cal{P}$ is given by \eq{solution}.

Different mechanisms of the magnetic field generation
in the early universe give the following main phenomenological
formula
\be[rms]
B = B_0 \Bigl (\frac{T}{T_0}\Bigr )^2\Bigl ( \frac{L_0}{l_H}\Bigr )^p,
\ee
where we put the maximal scale is chosen as the horizon
scale
\footnote{As explained in ref. \cite{EnqvistRezSemikoz},
this follows from the definition of the parameter $\lambda$ for the
initial $\delta$-correlator in \eq{correlator}.}
$L = l_H(t)$.

Note that if we put in \eq{rms} $B_0\sim T_0^2$
(as in Vachaspati's mechanism, where $B_0 \sim T_{EW}^2$
\cite{Vachaspati}) the r.m.s. field in \eq{solution}
depends only on the temperature of the universe $T$
($T \simeq T_{QCD}$ in \eq{BBN})
and on two parameters describing the geometry of the r.m.f.
geometry:
(a) the scaling parameter $p$, and
(b) the size of random domain $L_0^{min}= a(MeV/T)$.

Using this equation we get the following constraint
\footnote{Clearly in this approximation \eq{bbnconstraint}
leads to the same constraint on the transition magnetic moment
$\mu =\mu_{as}$ as obtained previously using kinetic theory
for the case of Dirac neutrino diagonal magnetic moment
\cite{EnqvistRezSemikoz}.}
\be[BBN]
\mu \lsim \frac{1.7\times 10^{-21}\mu_B}
{(L_0^{min})^{p + 1/2} \Gamma_W^{1/2}H^p},
\ee
where $L_0^{min}$ is the minimal scale of the random
magnetic field domains. Due to the copious production of
left-handed neutrinos e.g. from pion decays after
the quark hadron phase transition temperature
$T \simeq T_{QCD}  \simeq 200$ MeV this
QCD temperature plays a
crucial role in the estimate of the nucleosynthesis
constraint. As a result all parameters in \eq{BBN} are
evaluated at $T \simeq T_{QCD}  \simeq 200$ MeV.

The allowed region of transition
moments  $\mu_{\nu}$ we have obtained from nucleosynthesis
is shown in Fig. 1 as a function of the scale parameter $p$
and domain size $L_0$.

There is an independent bound from primordial nucleosynthesis
on the strength of the random magnetic
field. Indeed, it has been shown \cite{Schramm1} that a too
large r.m.f. strength would enhance the rates for the relevant
weak processes at $T_{NS} \sim 0.1$ MeV, resulting in additional
helium production. This leads to  $B\lsim 3\times 10^{10}$ Gauss
\cite{Schramm1} which implies, at the present time \cite{EnqvistRezSemikoz},
\be[strengthnow]
B_{NS} \lsim 1.8 \times 2.4^p \times 10^{-7 -11p} \rm{Gauss}
\ee
Clearly, all values of the parameter $p$ between $p = 0$
(corresponding to uniform field) and  $p = 3/2$
(corresponding to  3-dimensional elementary cells)
can obey both nucleosynthesis limits on
magnetic transition moments \eq{BBN}
and on r.m.f. strength, \eq{strengthnow}.

It is important to stress that, for a wide choice
of $p$ and $L_0$, our nucleosynthesis constraint on
the active sterile transition
moments  is substantially stronger
than the astrophysical one
from supernova 1987A \cite{BarbieriMohapatra}.

Until now our arguments are quite general. Now we
show that our constraints are consistent with the
hypothesis of a primordial origin of the observed
galactic magnetic fields \cite{Vachaspati,Olesen}
which would require additional restrictions on the
parameters $p$ and $L_0$, resulting in a correspondingly
stronger limit on the neutrino transition magnetic
moments.

This would require $L_0^{min} \gsim 10^3(MeV/T)$ cm
and $p \leq 1$.
The first restriction comes from requiring that
primordial fields survive against ohmical dissipation
after recombination time and could seed the
observed galactic magnetic field \cite{Schramm1}.
On the other hand, the limit  $p \leq 1$ follows from
\eq{strengthnow} and the lower dynamo theory astrophysical
bound on the strength of the galactic magnetic field
$B_{seed} (T_{now}) \gsim 10^{-18}$ Gauss \cite{Shukurov}.
The corresponding region is hatched in Fig. 1.
This corresponds to an upper limit on the neutrino
transition moment $\mu \lsim 10^{-16}\mu_B$ (left
corner of hatched region for $p =1$).
Moreover, in order to be a seed field for the
Milky Way or Andromeda, one would require
$B_{seed}(T_{now}) \gsim 10^{-13}$ Gauss \cite{Shukurov}
which corresponds to $p \lsim 0.5$ \cite{Olesen93}, leading
to a much tighter bound $\mu \lsim 10^{-19} \mu_B$.

\section{Supernova bounds}

Now let us move to the case of a supernova with magnetic field.
If this magnetic field is generated after collapse it
could be viewed as a random superposition of many small
dipoles of size $L_0 \sim 1$ km \cite{Dunkan}.
One can show that this hypothesis is consistent,
even for large magnetic field strengths,
with the non observation of gamma radiation from
SN 1987A, if the magnetic field diffusion time
\cite{Dunkan}
\be[Bdiffusiont]
t_{diff} \simeq 10^2 (B_0/10^9 G)^2 sec
\ee
does not exceed the time necessary for the diffusion
of the X-rays through the supernova mantle. This leads to an upper
limit on the seed field $B_0$, $B_0 \lsim 10^{12} G$.
Thus, as long as this is fulfilled, the random magnetic field
could well influence the SN 1987A neutrino burst without
being observable at present via X-ray emission
\footnote{
Alternatively, for the case of neutron
stars the assumption of random magnetic field domains
is consistent with their observed magnetic fields in
the limit where the small domains merge together to
form larger ones due to kinematic evolution of
the random magnetic field energy transferred
down the spectra $E(k) \sim  k^{-5/3}$ \cite{Zeldo}
with the characteristic time $ \tau (L_0) \sim  L_0^{2/3}$
}.

As we will show below, one can derive constraints
on the active to sterile Majorana neutrino magnetic
moments which are so stringent that they become relevant
even for values of the r.m.s field strength
as low as $B \lsim 10^{9}$ Gauss which could be
quite reasonable from the astrophysical point of view.

First, let us consider the strong small-scale
random magnetic fields generated
in a supernova as in \cite{Dunkan}.
For this case we can use the requirement that
in the non-trapping regime the
sterile neutrino can be emitted from anywhere
inside the stellar core with
 a rate \cite{sergio}
\be[volumelosses]
\frac{dQ}{dt} \simeq
2.8 \times 10^{55} \cal{P}  \rho_{14} E_{100}^{2} \frac{J}{s}
\lsim 10^{46}\frac{J}{s},
\ee
where $\rho_{14}$ is the core density in units of $10^{14}$ g/$cm^{3}$
and $E_{100}$ is the neutrino energy in units of 100 MeV. In
the following we will set these parameters to unity.
The last bound corresponds to the maximum observed
integrated neutrino luminosity. For instance, for the case
of SN1987A, this is $\sim 10^{46}J$.

Substituting our solution \eq{solution} to the luminosity bound
in \eq{volumelosses} we obtain the new astrophysical constraint
on the neutrino transition magnetic moment,
\be[supernova]
\mu \lsim \frac{10^{-8}\mu_B}{(B/1G)\sqrt{(L_0/1km)
\times (R_{core}/10km)}},
\ee
Thus one sees that, even for a reasonable r.m.s.
field $B \gsim 10^8$ $G$, this gives a more stringent
bound than obtained from SN1987A \cite{BarbieriMohapatra}.
It is also a more stringent bound than obtained above from
nucleosynthesis in the presence of a primordial r.m.f.
generated at the electroweak phase transition.

Note that the validity of \eq{supernova} requires averaging
over small random r.m.f. distributions, so that the
domain size $L_0$ must be much less than the core
radius $R_{core}$ and this, in turn, much less
than the damping length $\Gamma_{\perp}^{-1}$.
Indeed one can rewrite \eq{supernova} in
such a way that both of the conditions
\beq
L_0  \ll R_{core} \ll \Gamma_{\perp}^{-1} =
[(8/3)(\mu B)^2L_0]^{-1}
\eeq
are verified, provided $ L_0  \ll R_{core}$.
The bound shown in Fig. 2 corresponds to
$L_{0} \sim 10$ meters in \eq{supernova}.

\section{Discussion and conclusions}
\vskip 1cm

We have derived bounds on transition
neutrino magnetic moments connecting active to sterile
Majorana neutrinos $\nu_{a}$ and $\nu_{s}$ for
neutrinos which propagate in the presence of
random magnetic fields. We treated both the
constraints that follow from nucleosynthesis
as well as those that follow from supernova energy
loss arguments. In both cases our constraints
are typically much stronger than in the case
without random magnetic fields. The allowed
region of parameters corresponding to each case
are shown in Fig. 1 and Fig. 2. In the nucleosynthesis
case our bounds are consistent with the hypothesis of
seeding of galactic magnetic fields by primordial fields
generated at the electroweak transition, and
surviving past the re-combination epoch.
In the supernova case, even for a reasonable r.m.s.
field $B \gsim 10^8$ $G$, our constraint is more stringent
than obtained from SN1987A. It is also a more stringent
than the nucleosynthesis bound obtained in this paper
in the presence of a primordial r.m.f.
generated at the electroweak phase transition.

Our results were derived from a simple
Schr\"{o}dinger-type evolution equation
ignoring neutrino interactions. Our results
were derived for the general case of Majorana
neutrino transition magnetic moments, but in the
approximation of negligible mixing.
They are identical to the ones derived from
nucleosynthesis in ref. \cite{EnqvistRezSemikoz}
for the case of diagonal Dirac neutrino magnetic moments.
This suggests that kinetic theory effects
neglected here are not too important and
should encourage one to perform a complete
study in which the mixing between the two Majorana
neutrino species is taken into account. We plan to
come back to this question elsewhere.
\vfill
{\bf Acknowledgements:}
This work was supported by DGICYT under grant numbers
PB92-0084 and SAB94-0325. S. P. was supported by a
fellowship from Generalitat Valenciana. We thank Sasha
Dolgov for discussions.

\vskip 1truecm

\newpage
\begin{center}
\vskip 5cm
{\bf Figure Captions}
\end{center}
\vskip2cm

{\bf Fig.1.} \\

Constraints on active to sterile neutrino
transition magnetic moments derived
from early universe nucleosynthesis in the
presence of a strong random magnetic field.
This limit is given
as a function of the domain size and field geometry
parameter p, explained in the text.
The region corresponding to the seeding hypothesis of galactic
magnetic fields is indicated by the hatching.

\vskip 1cm

{\bf Fig.2.} \\

Constraints on active to sterile neutrino
transition magnetic moments derived
from supernova cooling arguments.
This limit is given as a function of
the random magnetic field strength
for typical values of neutrino energy,
core density and size, and lepton abundances.
Note however that, for the dynamo mechanism
suggested in ref. \cite{Dunkan} the seed field $B_0$ should be larger than
$B_0 \gsim 2 \times 10^8 G$.


\newpage

\begin{thebibliography}{99}

\bibitem{Zeldo}
Ya. Zeldovich, A. A. Ruzmaikin and D. D. Sokoloff,
Magnetic Fields in Astrophysics, Mc Graw Hill, 1983;
E. Parker, Cosmological Magnetic Fields,
Oxford Univ. Press, 1979;
 A. A. Ruzmaikin, A. A. Shukurov and D. D. Sokoloff,
Magnetic Fields of Galaxies, Kluver, 1988.

\bibitem{granadasol}
T. Kirsten, \pl{B314}{93}{445};
Kamiokande, SAGE, GALLEX collaboration talks
at {\sl XV Int. Conference on Neutrino Physics
and Astrophysics}, \nps{31}{93}{105-124};
 J. R.~Davis in {\sl Proceedings of the 21th
International Cosmic Ray Conference,  Vol. 12},
ed.\  R.~J. Protheroe (University of Adelaide Press, 1990) p. 293.

\bibitem{atm}
Kamiokande collaboration, \pl{B205}{88}{416},
\pl{B280}{92}{146} and \pl{B283}{92}{446} ;
IMB collaboration, \pr{D46}{92}{3720};
see also the proceedings of {\it Int. Workshop on
\nm/\ne problem in atmospheric neutrinos} ed. V. Berezinsky
and G Fiorentini, Gran Sasso, 1993.

\bibitem{cobe}
G.~F. Smoot et~al., \apj{396}{92}{L1-L5}.

\bibitem{cobe2}
E.L.~Wright et al., \apj{396}{92}{L13};
M.~Davis, F.J.~Summers, and D.~Schagel, \nat{359}{92}{393};
A.N.~Taylor and M.~Rowan-Robinson, \ib{359}{92}{396};
R.K.~Schaefer and Q.~Shafi, \nat{359}{92}{199};
J.A.~Holtzman and J.R.~Primack, \apj{405}{93}{428};
A.~Klypin et al., \apj{416}{93}{1}.

\bibitem{DARK92}
J.~T. Peltoniemi, D.~Tommasini, and J. W. F. Valle,
\pl{B298}{93}{383}.

\bibitem{DARK92B}
J.~T. Peltoniemi, and J. W. F. Valle, \np{B406}{93}{409-422};
D.O.~Caldwell and R.N.~Mohapatra, \pr{D48}{93}{3259}.

\bibitem{Schramm}
T.Walker, G.Steigman, D.N.Schramm, K.Olive, and H.Kang,
\apj{376}{91}{51}.

\bibitem{Enqvist}
K. Enqvist, K. Kainulainen and M. J.  Thomson,
\np{B373}{92}{498};
K.Kainulainen, \pl{B224}{90}{191};
R.Barbieri and A.Dolgov, \pl{B237}{90}{440};
\np{B349}{91}{742};
J. Cline, \prl{68}{92}{3137}.

\bibitem{SemikozValle}
V. Semikoz and J. W. F. Valle, \np{B425}{94}{651}.

\bibitem{Maalampi}
K. Kainulainen, J. Maalampi and J.T. Peltoniemi, \np{B358}{91}{435};
G. Raffelt and G. Sigl, \ap{1}{93}{165}.

\bibitem{sergio}
S. Pastor, V. Semikoz, J.W.F.Valle, \ap{3}{95}{87}.

\bibitem{Dunkan}
C.Thomson and R.C.Dunkan, \apj{408}{93}{194}.

\bibitem{BFD}
J. Schechter and  J. W. F. Valle, \pr{D24}{81}{1883};
\pr{D25}{82}{283}.

\bibitem{LAM}
C.-S. Lim and W.J. Marciano, \pr{D37}{88 }{1368};\\
E. Kh. Akhmedov, \pl{B213}{88}{64}.

\bibitem{BarbieriMohapatra}
I. Goldman et al., \prl{60}{88}{1789};\\
J.M. Lattimer and J. Cooperstein, \prl{61}{88}{23};\\
R. Barbieri and R.N. Mohapatra, \prl{61}{88}{27};\\
M. B. Voloshin, \pl{B209}{88}{360}.

\bibitem{Sergio95}
S. Pastor, V. Semikoz and J.W.F. Valle, (in preparation).

\bibitem{Notzold}
D.N\"{o}tzold, G.Raffelt, \np{B307}{88}{924}.

\bibitem{EnqvistRezSemikoz}
K.Enqvist, A.Rez and V.Semikoz, \np{B436}{95}{49}.

\bibitem{Semikoz}
V. Semikoz, \pr{D48}{93}{5264}.

\bibitem{BalantekinLoreti}
 F.N. Loreti and A.B. Balantekin, \pr{D50}{94}{4762}.

\bibitem{Vachaspati}
T.Vachaspati, \pl{B265}{91}{258}.

\bibitem{Schramm1}
B. Cheng, D. N.  Schramm, J. W. Truran, \pr{D49}{94}{5006};
D. Grasso and H. Rubinstein, \ap{3}{95}{87}.

\bibitem{Olesen}
P.Olesen, \pl{B281}{92}{300}.

\bibitem{Shukurov}
K.Enqvist, V.Semikoz, A.Shukurov and D.Sokoloff, \pr{D48}{93}{4557}.

\bibitem{Olesen93}
K.Enqvist and P.Olesen, \pl{B319}{93}{178}.

\end{thebibliography}
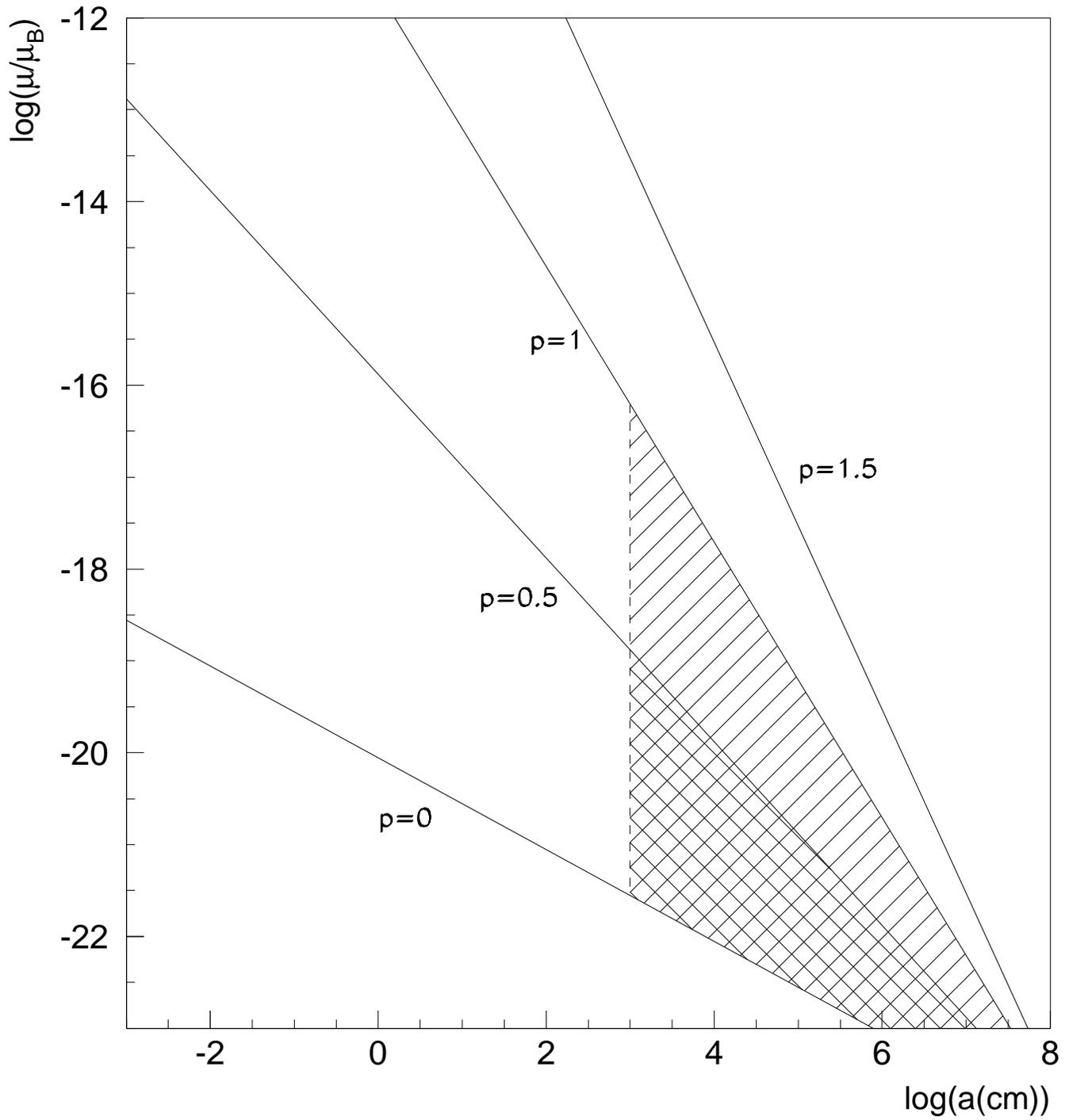
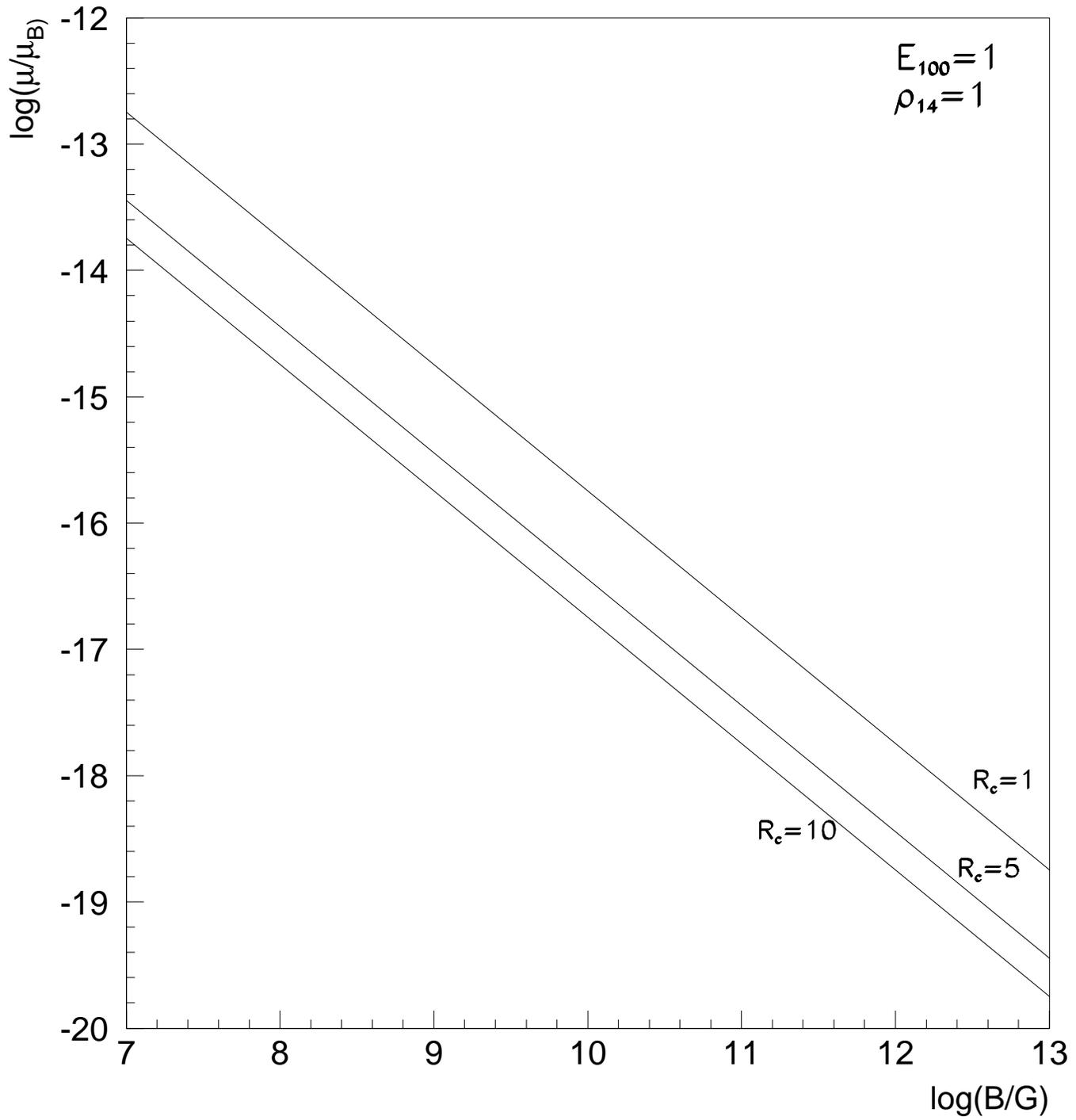
\end{document}